\documentclass[amsmath,pra,twocolumn,showpacs]{revtex4}
\usepackage{graphics}

\begin{document}     

\title{State-insensitive trapping and guiding of cesium atoms using a two-color evanescent field around a subwavelength-diameter fiber}
\author{ Fam Le Kien,$^{1,*}$ V. I. Balykin,$^{1,2}$ and K. Hakuta$^{1}$} 
\affiliation{
$^1$Department of Applied Physics and Chemistry, 
University of Electro-Communications, Chofu, Tokyo 182-8585, Japan\\
$^2$Institute of Spectroscopy, Troitsk, Moscow Region, 142092, Russia}
\date{\today}

\begin{abstract}
We calculate the optical potentials, i.e. the light shifts, of the ground and excited states of atomic cesium in a two-color evanescent field around a subwavelength-diameter fiber. We show that the light shifts  of the $6S_{1/2}\leftrightarrow 6P_{3/2}$  transitions can be minimized by tuning one trapping light to around $934.5$ nm in wavelength (central red-detuned magic wavelength) and the other light to around  $685.5$ nm in wavelength (central blue-detuned magic wavelength). The simultaneous use of the red- and blue-detuned magic wavelengths allows state-insensitive two-color trapping and guiding of cesium atoms along the thin fiber. Our results can be used to efficiently load a two-color dipole trap by cesium atoms from
a magneto-optical trap and to perform continuous observations.
\end{abstract}

\pacs{32.80.Pj,32.80.Lg,03.75.Be,03.65.Ge}
\maketitle

\section{Introduction}

There is currently a strong interest for the manipulation of individual neutral atoms in microscopic (at subwavelength size) optical dipole traps \cite{Schlosser,Kuhr}. The ability to manipulate atoms individually may open a way to controlled engineering of the quantum states of small sets of trapped particles, in order to encode and process information at the quantum level \cite{Sackett}. A red-detuned optical dipole trap works as follows: The ac Stark shift induced by a red-detuned trapping light lowers the ground-state energy of the atom proportionally to the local light intensity \cite{dipoleforce}. The spatial dependence of the atomic potential energy is therefore equivalent to the spatial dependence of the light intensity. The atom has the lowest potential energy in the region of maximum intensity and can therefore be trapped there. For very large detuning, typically several nanometers, the photon scattering rate becomes so low that the optical trapping potential is truly conservative. The optical dipole trap is usually loaded by precooled atoms from a magneto-optical trap (MOT). There are two schemes of \textit{microscopic} optical dipole traps to store individual atoms. The first one is based on a \textit{strongly focused} single Gaussian laser beam  \cite{Schlosser}. The second one is based on the use of a \textit{standing wave} of two counterpropagating laser beams with equal intensities and optical frequencies \cite{Kuhr}. In both schemes, the size of the trapping region is less than the light wavelength.

Recently, we proposed a new method for \textit{microscopic} trapping of individual atoms 
\cite{paper1,twocolors}.
The method is based on the use of a subwavelength-diameter silica fiber with a single (red-detuned) light beam \cite{paper1} or with two (red- and blue-detuned) light beams 
\cite{twocolors} launched into it. 
The light waves decay away from the fiber wall and produce optical potentials for  neutral atoms. 
The red-detuned light wave produces an attractive potential. 
To sustain stable trapping, the atoms must be kept away from the fiber wall. This can be achieved by the centrifugal potential barrier in the one-color scheme \cite{paper1}  or by the repulsive optical potential from the blue-detuned laser beam in the two-color scheme \cite{twocolors,Ovchinnikov}. 
The atom trapping and guiding occur outside the fiber. We have shown that stable trapping and guiding can be achieved only when the fiber diameter is smaller than the light wavelength.
The great advantages of our scheme are (a) localization of atoms to a subwavelength region, (b) high efficiency to detect individual atoms, (c) high accessibility to the trapped atoms, and (d) achievement of strong coupling between light and matter \cite{Domokos}. 

Due to the conservative character of optical potentials, the loading of atoms into  dipole
traps requires the use of friction forces, which can be provided by the Doppler cooling mechanism in  MOTs \cite{dipoleforce,coolingbook}. The loading of a dipole trap from a MOT is a dynamical process rich in physics \cite{Kuppens}.  
A major obstacle to the efficient loading of a dipole trap is that the excited electronic states generally experience positive ac Stark shifts of comparable magnitude to the negative shift of the ground state. 
These light shifts with opposite signs change the resonant frequencies of the atoms in the dipole trap. When the spatial gradient of the trapping fields is high, the resonant frequencies of the atoms vary substantially with their positions  within the dipole trap. 
This effectively introduces a strong spatial dependence of the detuning between the atomic cooling transition and the MOT fields. The detuning may become large or positive. This prevents the efficient Doppler cooling.  As a result, the Doppler cooling is generally  incompatible with the dipole trapping. The extremely small volume of a microscopic dipole trap makes the problem even worse: the time-averaged number of atoms in a microscopic dipole trap (loaded from a MOT) can be less than one. 

Katori \textit{et al.} have proposed and demonstrated a trap loading scheme that helps overcome the above obstacle \cite{Katori}. The idea is to use an appropriate wavelength of the trapping laser light (called the magic wavelength) at which (due to the specific multilevel structure of a real atom) the Stark shifts of the lower and upper levels of the cooling transition have the same value and the same sign. It has been demonstrated for strontium atoms that the choice of a magic wavelength for a red-detuned far-off-resonance trap allows simultaneous Doppler cooling and dipole trapping and hence enables high loading efficiency of magneto-optically trapped atoms into the optical dipole trap \cite{Katori}. 

The spatial dependence of the atomic resonant frequencies also leads 
to additional complications in the monitoring (probing) processes. Kimble \textit{et al.} have recently demonstrated for cesium atoms that the choice of a magic wavelength for a red-detuned far-off-resonance trap allows state-insensitive trapping and continuous observation of trapped atoms \cite{Kimble}.
  
In this paper, we address the problem of minimizing the spatial dependence of the light shifts of the atomic transitions in a two-color dipole trap. 
We show that the light shifts  of the $6S_{1/2}\leftrightarrow 6P_{3/2}$ cesium  transitions can be minimized by tuning one trapping light to around  $934.5$ nm in wavelength (central red-detuned magic wavelength) and the other light to around $685.5$ nm in wavelength (central blue-detuned magic wavelength).
We calculate the optical potentials of cesium atoms in the ground and excited states in a two-color evanescent field around a subwavelength-diameter fiber. 
We show the possibility of state-insensitive two-color trapping and guiding.
  
Before we proceed, we note that, due to recent developments in taper fiber technology, thin fibers can be produced with diameters down to 50 nm \cite{Mazur'sNature,Birks}. 
Thin fiber structures can be used as building blocks in future atom and photonic micro- and nano-devices.
 
The paper is organized as follows. In Sec.\ \ref{sec:theory} we review the general theory and present the basic equations for light shifts and polarizabilities of atomic states. 
In Sec.\ \ref{sec:numerical} we calculate the dynamic polarizabilities of the $6S_{1/2}$ ground state and  the $6P_{3/2}$ excited state of atomic cesium. In Sec.\ \ref{sec:Stark} we calculate the light shifts of these  states in a two-color evanescent field around a subwavelength-diameter fiber. 
Our conclusions are given in Sec.~\ref{sec:summary}.

\section{General theory of light shifts of atomic hyperfine levels}
\label{sec:theory}

We consider the interaction between an atom in a fine-structure state $|n\rangle\equiv|nL_{J}\rangle$ and an external electric field $\mathbf{E}$.
The combined Hamiltonian of the hyperfine interaction and the Stark effect is
\begin{equation}
H=V_{\mathrm{hfs}}+V_{EE},
\label{8}
\end{equation}
where the operator $V_{\mathrm{hfs}}$ describes the hyperfine 
structure and the operator $V_{EE}$ describes the field-induced shifts of energy levels. 

The hyperfine interaction operator $V_{\mathrm{hfs}}$ is given by \cite{Schwartz}
\begin{equation}
V_{\mathrm{hfs}}=\hbar A\,\mathbf{I}\cdot\mathbf{J}+\hbar B\frac{6(\mathbf{I}\cdot\mathbf{J})^2+3\mathbf{I}\cdot\mathbf{J}-2I(I+1)J(J+1)}{2I(2I-1)2J(2J-1)}.
\label{6}
\end{equation}
Here $\mathbf{J}$ is the operator for the total electronic angular momentum,
$\mathbf{I}$ is the operator for the nuclear spin, and $A$ and $B$ are the hyperfine-structure (hfs) constants. For the level $6P_{3/2}$ of cesium ($I=7/2$ and $J=3/2$), we have \cite{coolingbook} $A/2\pi=50.34$ MHz and $B/2\pi=-0.38$ MHz. 

The Stark operator $V_{EE}$ is, in the second-order perturbation theory, given by \cite{Schmieder}
\begin{equation}
V_{EE}=-\frac{1}{2}\alpha_0 E^2-\frac{1}{2}\alpha_2 Q E^2.
\label{1}
\end{equation}
Here $\alpha_0$ and $\alpha_2$ are the scalar and tensor polarizabilities, respectively. The scalar polarizability $\alpha_0$ shifts all hyperfine and magnetic sublevels equally. The tensor polarizability $\alpha_2$ mixes the hyperfine and magnetic sublevels through the operator 
\cite{Schmieder}
\begin{equation}
Q=\frac{3 (\mathbf{u}\cdot\mathbf{J})^2 -J(J+1)}{J(2J-1)}.
\label{2}
\end{equation}
Here $\mathbf{u}=\mathbf{E}/E$ is the unit vector in the field direction. 

Due to the hfs interaction, the total electronic angular momentum $\mathbf{J}$ is not conserved. However, in the absence of the field, the total angular momentum of the atom,  
described by the operator $\mathbf{F}=\mathbf{J}+\mathbf{I}$, is conserved. 
In the basis of hfs states $|FM_F\rangle$,
the operator $V_{\mathrm{hfs}}$ is diagonal. Its nonzero matrix elements are  
\begin{equation}\label{7}
\begin{split}
&\langle FM_F|V_{\mathrm{hfs}}|FM_F\rangle \\
&=\frac{1}{2}\hbar AK+\hbar B\frac{\frac{3}{2}K(K+1)-2I(I+1)J(J+1)}{2I(2I-1)2J(2J-1)},
\end{split}
\end{equation}
where $K=F(F+1)-I(I+1)-J(J+1)$.

The matrix elements of $Q$ between two hfs states are given by \cite{Schmieder}
\begin{eqnarray}
\lefteqn{\langle FM_F|Q|F^\prime M^\prime_{F}\rangle}
\nonumber\\&&
=\sqrt{\frac{15}{2}}\left[\frac{(J+1)(2J+1)(2J+3)}{J(2J-1)} \right]^{1/2}
\nonumber\\&&\mbox{}\times
\sum_{q}\sum_{\mu=-1}^1\sum_{\mu^\prime=-1}^1
u_\mu u_{\mu^\prime }
\left(\begin{array}{ccc}1&2&1\\ \mu&-q&\mu^\prime \end{array}\right)
\nonumber\\&&\mbox{}\times
(-1)^{I+J+F-F^\prime -M_F}\sqrt{(2F+1)(2F^\prime +1)}
\nonumber\\&&\mbox{}\times
\left(\begin{array}{ccc}F&2&F^\prime \\ M_F&q&-M_F^\prime \end{array}\right)
\left\{\begin{array}{ccc}F&2&F^\prime \\ J&I&J\end{array}\right\}.
\label{3}
\end{eqnarray}
Here $u_{-1}=(u_x-iu_y)/\sqrt{2}$, $u_0=u_z$, and $u_{1}=-(u_x+iu_y)/\sqrt{2}$ are the spherical tensor components of the field-direction vector $\mathbf{u}$. The three-$j$ symbols in eq.~(\ref{3}) require the two conditions $q=\mu+\mu^\prime=M_F^\prime-M_F$. The second three-$j$ symbol and the six-$j$ symbol in eq.~(\ref{3}) make $Q$ not diagonal in $F$.

Equations (\ref{2}) and (\ref{3}) are valid for an arbitrary orientation of the electric field. In a particular case where the electric field is aligned along the quantization axis $z$, i.e., $\mathbf{E}=E \hat{\mathbf{z}}$, eq. (\ref{2}) reduces to the form
\begin{equation}
Q=\frac{3J_z^2 -J(J+1)}{J(2J-1)},
\label{2a}
\end{equation}
which is diagonal in $J$ and $M_J$.
Most of the previous works on the Stark effect in atomic excited states were devoted to this case  \cite{Schmieder,Khadjavi,Schmieder71}. 
 
In the absence of the hfs interaction, the Stark shift of a fine-structure magnetic sublevel $|JM_J\rangle$ can be written as 
$\Delta E_{M_J}=-(1/2)\alpha(M_J) E^2$, with the polarizability 
\begin{equation}
\alpha(M_J)=\alpha_0+\alpha_2\frac{3M_J^2 -J(J+1)}{J(2J-1)}.
\label{2b}
\end{equation}
In particular, for $J=3/2$, we have $\alpha(M_J=\pm3/2)=\alpha_0+\alpha_2$ and $\alpha(M_J=\pm1/2)=\alpha_0-\alpha_2$. 

The presence of the hfs interaction dictates the use of the hfs basis  $\{|FM_F\rangle\}$. 
In this basis, the particular special form (\ref{2a}) of the operator $Q$ is  diagonal in 
$M_F$ but not in $F$. The nonzero matrix elements of this operator are \cite{Schmieder}
\begin{eqnarray}
\lefteqn{\langle FM_F|Q|F^\prime M_{F}\rangle=\left[\frac{(J+1)(2J+1)(2J+3)}{J(2J-1)} \right]^{1/2}}
\nonumber\\&&\mbox{}\times
(-1)^{I+J+F-F^\prime -M_F}\sqrt{(2F+1)(2F^\prime +1)}
\nonumber\\&&\mbox{}\times
\left(\begin{array}{ccc}F&2&F^\prime \\ M_F&0&-M_F \end{array}\right)
\left\{\begin{array}{ccc}F&2&F^\prime \\ J&I&J\end{array}\right\}.
\label{3a}
\end{eqnarray}
For each fixed value of $M_F$, there is a matrix with rows and columns labeled by $F$ and $F^\prime$,
respectively. In particular, for $M_F=\pm F_{\mathrm{max}}$, where $F_{\mathrm{max}}\equiv J+I$,
the matrix $\langle FM_F|Q|F^\prime M_{F}\rangle$ reduces to $1\times 1$. 
Hence, for the hfs states  with the maximum values of $F$ and $|M_F|$, i.e. the states  
$|F=F_{\mathrm{max}},M_F=\pm F_{\mathrm{max}}\rangle$, the Stark shift is 
$\Delta E_{M_\mathrm{max}}=-(1/2)\alpha_{M_\mathrm{max}} E^2$,  
where $\alpha_{M_\mathrm{max}}=\alpha_0+\alpha_2$. 

In this paper we study the case where the field interacting with the atom  is an optical field, that is, $\mathbf{E}=(\vec{\mathcal{E}}e^{-i\omega t}+\vec{\mathcal{E}}^\ast e^{i\omega t})/2$. 
Here $\vec{\mathcal{E}}$ is the complex envelope vector of the electric component of the light field. In this case, we have to remove fast optical oscillations from the right-hand side of 
eq.~(\ref{1}) by averaging it over an optical period. This procedure leads to the following expression for the operator of the dynamic Stark effect:
\begin{equation}\label{3b}
\begin{split}
V_{EE}=& -\frac{1}{4}\alpha_0 |\mathcal{E}|^2
-\frac{1}{4}\alpha_2\sum_{ FM_FF^\prime M^\prime_{F}}|FM_F\rangle\langle F^\prime M^\prime_{F}| \\
& \times\sum_{q\mu\mu^\prime}(-1)^{\mu^\prime}
\mathcal{E}_\mu \mathcal{E}_{-\mu^\prime}^\ast 
\left(\begin{array}{ccc}1&2&1\\ \mu&-q&\mu^\prime \end{array}\right)  \\
& \times\sqrt{\frac{15}{2}}\left[\frac{(J+1)(2J+1)(2J+3)}{J(2J-1)} \right]^{1/2}  \\
& \times(-1)^{I+J+F-F^\prime -M_F}\sqrt{(2F+1)(2F^\prime +1)}  \\
& \times\left(\begin{array}{ccc}F&2&F^\prime \\ M_F&q&-M_F^\prime \end{array}\right)
\left\{\begin{array}{ccc}F&2&F^\prime \\ J&I&J\end{array}\right\}.
\end{split}
\end{equation}
Here $\mathcal{E}_{-1}=(\mathcal{E}_x-i\mathcal{E}_y)/\sqrt{2}$, $\mathcal{E}_0=\mathcal{E}_z$, and $\mathcal{E}_{1}=-(\mathcal{E}_x+i\mathcal{E}_y)/\sqrt{2}$ are the spherical tensor components of the field envelope vector  $\vec{\mathcal{E}}$. In general, the Stark operator $V_{EE}$ is not diagonal in the hfs basis  $\{|FM_F\rangle\}$, and consequently neither is the Hamiltonian (\ref{8}). To find the shifts of the hfs sublevels, we must diagonalize this Hamiltonian. The Stark shift induced by an optical field is called the ac Stark shift or the light shift.

In the case of optical fields, in addition to the time averaging procedure, we also have to use the dynamic polarizability instead of the static one.
The static scalar  and tensor polarizabilities  of an atomic fine-structure state have been derived systematically \cite{Khadjavi}. The microscopic expression for the linear susceptibility, which is related to the dynamic polarizability, is well known in the literature \cite{Boyd}. 
Combining the previous  results \cite{Khadjavi,Boyd}, we  write 
the dynamic scalar polarizability $\alpha_0$ and the dynamic tensor polarizability $\alpha_2$ of an atomic fine-structure state $|n\rangle$ as
\begin{eqnarray}
\alpha_0&=&\frac{2/\hbar}{3(2J+1)}\sum_{n^\prime }\langle n^\prime \| D\| n\rangle^2
\nonumber\\&&\mbox{}\times
\frac{\omega_{n^\prime n}(\omega_{n^\prime n}^2-\omega^2+\gamma_{n^\prime n}^2/4)}{(\omega_{n^\prime n}^2-\omega^2+\gamma_{n^\prime n}^2/4)^2+\gamma_{n^\prime n}^2\omega^2}
\label{4}
\end{eqnarray}
and
\begin{eqnarray}
\alpha_2&=&\frac{4}{\hbar}\left(\frac{5J(2J-1)}{6(J+1)(2J+1)(2J+3)}\right)^{1/2}
\nonumber\\&&\mbox{}
\times
\sum_{n^\prime }(-1)^{J+J^\prime }\left\{\begin{array}{ccc}J&1&J^\prime \\1&J&2\end{array}\right\}\langle n^\prime \| D\| n\rangle^2
\nonumber\\&&\mbox{}
\times
\frac{\omega_{n^\prime n}(\omega_{n^\prime n}^2-\omega^2+\gamma_{n^\prime n}^2/4)}{(\omega_{n^\prime n}^2-\omega^2+\gamma_{n^\prime n}^2/4)^2+\gamma_{n^\prime n}^2\omega^2},
\label{5}
\end{eqnarray}
respectively. Here $\omega$ is the frequency of the light field, $\langle n^\prime \| D\| n\rangle$ is the reduced electric dipole matrix element for the transition between the fine-structure states $|n^\prime \rangle$ and $|n\rangle$, $\omega_{n^\prime n}=\omega_{n^\prime }-\omega_n$ is the transition frequency, and $\gamma_{n^\prime n}=\gamma_{n^\prime }+\gamma_n$ is  the linewidth (twice the dephasing rate) and is given as the sum of the population decay rates $\gamma_{n^\prime }$ and $\gamma_n$ of the states. We note that, when $\omega$ is several linewidths off resonance with the corresponding transition, the effect of $\gamma_{n^\prime n}$ in the above formulae can be neglected. We also note that, for the ground states of alkali-metal atoms, which correspond to $J=1/2$, 
the tensor polarizability is vanishing, that is, $\alpha_2=0$.

\section{Numerical results for cesium atoms in a linearly polarized plane-wave light field}
\label{sec:numerical}

\subsection{Dynamic polarizabilities of the $6S_{1/2}$ ground state 
and the $6P_{3/2}$ excited state}

The polarizabilities of the ground and excited states of atomic cesium
have been calculated in a large number of works. However, most of the previous calculations were devoted to the static limit \cite{Schmieder,Khadjavi,Safronova}. 
Recently, in order to search for a red-detuned magic wavelength for a far-off-resonance trap, the light shifts of the ground and excited states of atomic cesium have been calculated \cite{Kimble}. However, the results for the dynamic polarizabilities have not been explicitly provided. 

To search for red- and blue-detuned magic wavelengths,
we first calculate the polarizabilities of the ground and excited states of cesium as functions of the light wavelength $\lambda$, using eqs. (\ref{4}) and (\ref{5}).
The calculations for the polarizability of the $6S_{1/2}$ ground state incorporate the couplings 
$6S_{1/2}\leftrightarrow (6\text{--}11)P_{1/2,3/2}$.
The calculations for the scalar and tensor polarizabilities of the $6P_{3/2}$ excited state 
incorporate the couplings $6P_{3/2}\leftrightarrow (6\text{--}15)S_{1/2}$ and 
$6P_{3/2}\leftrightarrow (5\text{--}11)D_{3/2,5/2}$.
Relevant parameters are taken from a number of sources \cite{Safronova,Fabry,Moore,Theodosiou}. 

\begin{figure}[tbh]
\begin{center}
  \includegraphics{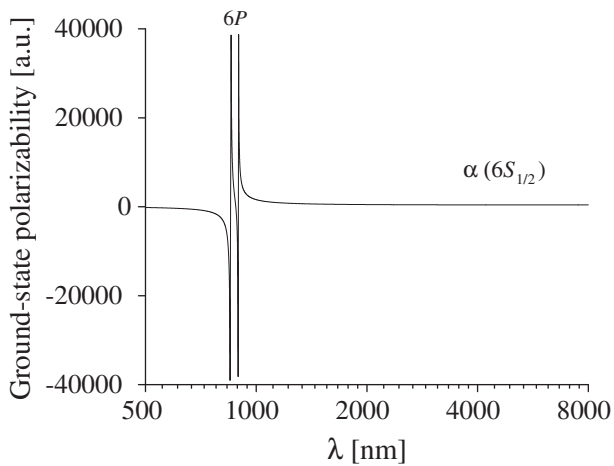}
 \end{center}
\caption{Polarizability $\alpha$  of the
ground state $6S_{1/2}$ in atomic cesium as a function of the light wavelength $\lambda$.}
\label{fig1}
\end{figure}

We plot in Fig. \ref{fig1} the dynamic polarizability $\alpha=\alpha_0$ of the ground state $6S_{1/2}$. As seen, in the region $\lambda>500$ nm, the profile of $\alpha$ has  two closely positioned resonances, corresponding to the transitions between 
$6S_{1/2}$ and $6P_{1/2}$ ($D_1$ line, wavelength 894 nm) and between 
$6S_{1/2}$ and $6P_{3/2}$ ($D_2$ line, wavelength 852 nm). The effects of the other transitions are not substantial in this wavelength region.

\begin{figure}[tbh]
\begin{center}
  \includegraphics{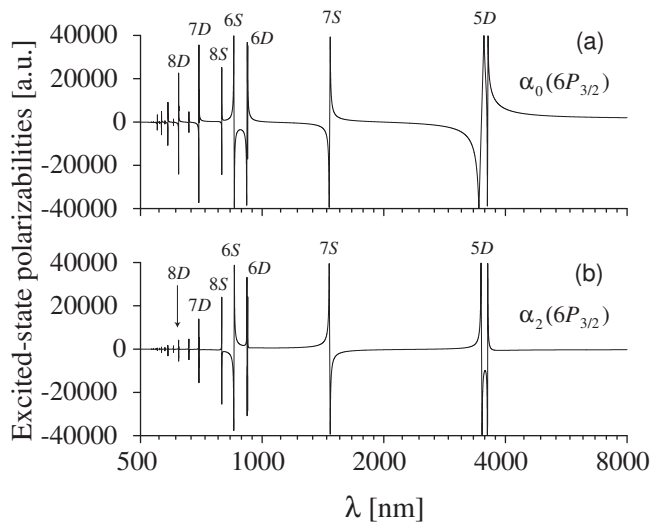}
 \end{center}
\caption{Scalar polarizability $\alpha_0$ (a) and tensor polarizability $\alpha_2$ (b) of the
excited state $6P_{3/2}$ in atomic cesium as functions of the light wavelength $\lambda$.}
\label{fig2}
\end{figure}

We plot in Fig. \ref{fig2} the scalar polarizability $\alpha_0$ and the tensor polarizability $\alpha_2$ for the excited state $6P_{3/2}$. The figure shows that both 
$\alpha_0$ and $\alpha_2$ have multiple resonances  in the region $\lambda>500$ nm. The most dominant resonances are due to the transitions from  $6P_{3/2}$ to (6--8)$S_{1/2}$ and (5--8)$D_{3/2,5/2}$.

\subsection{Blue- and red-detuned magic wavelengths}

\begin{figure*}
\begin{center}
  \includegraphics{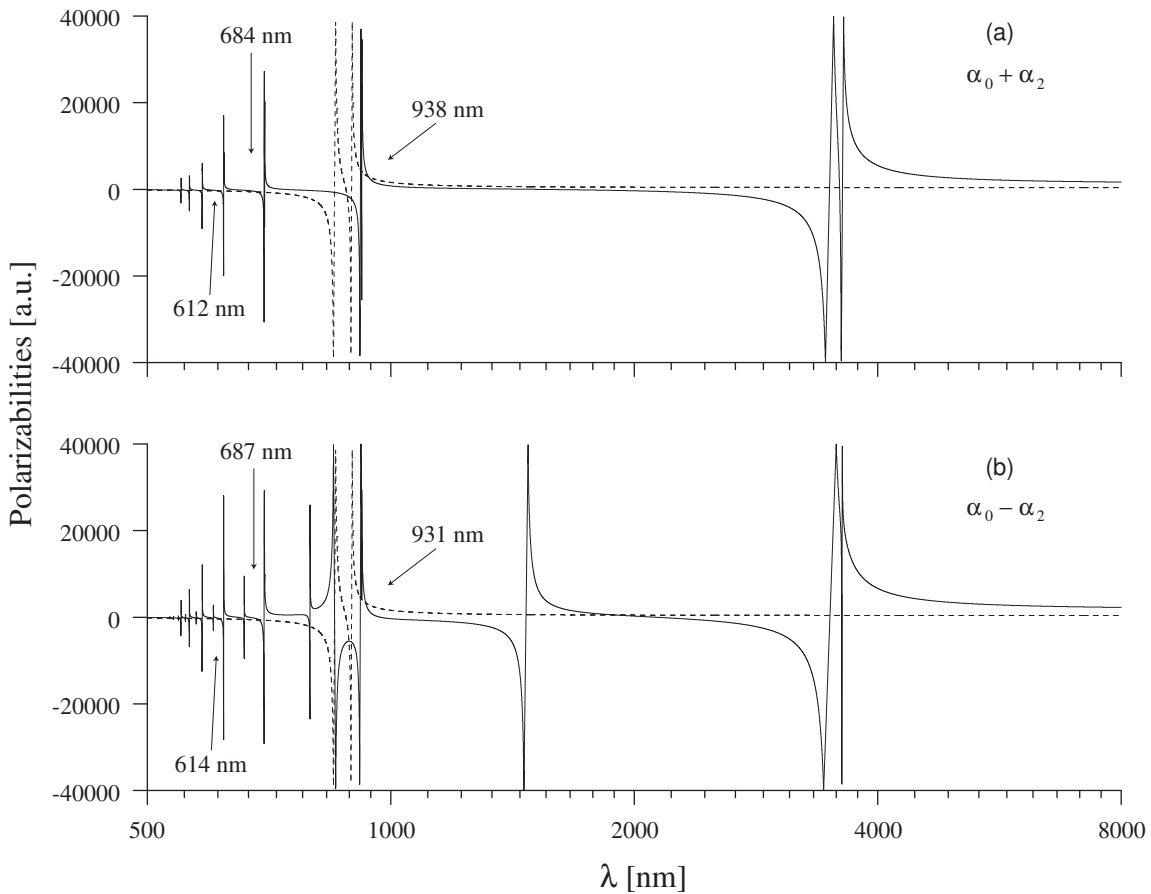}
 \end{center}
\caption{Comparison between the polarizabilities of the excited and ground states  of atomic cesium. The sum (a) and difference (b) of the scalar and tensor polarizabilities of the 
$6P_{3/2}$ excited state are shown as functions of the light wavelength by the solid lines. The polarizability of the $6S_{1/2}$ ground state is shown by the dashed lines.
}
\label{fig3}
\end{figure*}

We now search for magic wavelengths, at which the polarizabilities and consequently the light 
shifts of the relevant upper and lower states are almost equal, leading to the minimization of the shift of the atomic transition frequency \cite{Katori}. For this purpose,
we plot in Fig. \ref{fig3} the sum $\alpha_0+\alpha_2$ (a) and difference $\alpha_0-\alpha_2$ (b) of the scalar and tensor polarizabilities of the $6P_{3/2}$ excited state (solid lines) 
together with the 
polarizability of the $6S_{1/2}$ ground state (dashed lines). As mentioned in the previous section, the quantities $\alpha_0+\alpha_2$ and $\alpha_0-\alpha_2$ are the polarizabilities of the fine-structure magnetic sublevels with 
$M_J=\pm3/2$ and $M_J=\pm1/2$, respectively, in the case where the field is linearly polarized along the $z$ axis. Therefore, although the total polarizability of the $6P_{3/2}$ excited state is a tensor, the quantities $\alpha_0+\alpha_2$ and $\alpha_0-\alpha_2$ characterize the boundary magnitudes of the total polarizability.

As seen from Fig. \ref{fig3}, the sum $\alpha_0+\alpha_2$ and difference $\alpha_0-\alpha_2$ for the $6P_{3/2}$ state cross the polarizability of the $6S_{1/2}$ state
at slightly differing wavelengths of 938 nm and 931 nm, respectively. These crossing  points are red-detuned from the $D_1$ and $D_2$ resonance lines. 
They are spread around the central red-detuned magic wavelength  $\bar{\lambda}_R=934.5$ nm, in agreement with the recent result of McKeever \textit{et al.} for atomic cesium \cite{Kimble}. 

We observe in Fig. \ref{fig3} that, in addition to the crossings of  the ground- and excited-state polarizabilities on the red side of detuning, there are  several crossings on the blue side. The two blue-detuned crossings that are closest to the $D_1$ and $D_2$ resonance lines occur, in the case of $\alpha_0+\alpha_2$, at the wavelengths of 684 nm and 612 nm 
[see Fig. \ref{fig3}(a)] and, in the case of $\alpha_0-\alpha_2$, at the wavelengths of 687 nm and 614 nm [see Fig. \ref{fig3}(b)]. The difference between the positions of the $\alpha_0+\alpha_2$ and $\alpha_0-\alpha_2$ crossings is rather small. 
The first blue-detuned crossings are spread around the central magic wavelength  $\bar{\lambda}_{B}=685.5$ nm. The second blue-detuned crossings are spread around the central magic wavelength $\bar{\lambda}^\prime_B=613$ nm. 
Thus we can minimize the light shifts of the $6S_{1/2}\leftrightarrow 6P_{3/2}$ transitions of cesium atoms in a blue-detuned light field by tuning the light field 
to around an average wavelength $\bar{\lambda}_{B}=685.5$ nm or $\bar{\lambda}^\prime_B=613$ nm. 
Concerning the problem of trapping and guiding atoms around a subwavelength-diameter fiber, 
the first blue-detuned magic wavelength $\bar{\lambda}_{B}=685.5$ nm is more favorable than the second blue-detuned magic wavelength $\bar{\lambda}^\prime_B=613$ nm. 
One of the reasons is that the first wavelength is closer to the $D_1$ and $D_2$ resonance lines and hence leads to a larger coupling strength. In addition, the first wavelength satisfies better the single-mode fiber condition and gives a longer evanescent-wave penetration length. 
Therefore, we focus on the first blue-detuned magic wavelength but not on the second one. 

\begin{figure}
\begin{center}
  \includegraphics{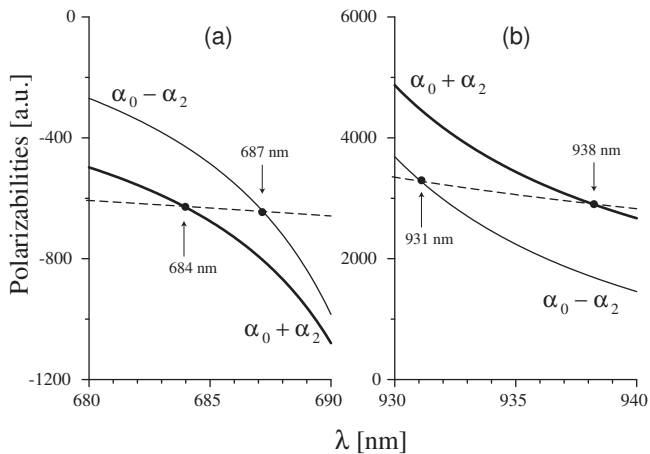}
 \end{center}
\caption{Blue-detuned (a) and red-detuned (b)  magic wavelengths for the 
$6S_{1/2}\leftrightarrow 6P_{3/2}$ cesium transition. The sum and difference of the scalar and tensor polarizabilities of the  $6P_{3/2}$ excited state are shown by the thick and thin solid lines, respectively. The polarizability of the $6S_{1/2}$ ground state is shown by the dashed lines.}
\label{fig4}
\end{figure}

We show in  Figs. \ref{fig4}(a) and \ref{fig4}(b) the polarizabilities of the $6S_{1/2}$ and 
$6P_{3/2}$ states in the vicinities of the central blue-detuned magic wavelength $\bar{\lambda}_{B}=685.5$ nm and the central red-detuned magic wavelength $\bar{\lambda}_R=934.5$ nm, respectively. As seen, around $\bar{\lambda}_{B}$ and $\bar{\lambda}_R$, the polarizabilities of the ground and excited states cross each other. The signs of the polarizabilities in the vicinities of $\bar{\lambda}_{B}$  and $\bar{\lambda}_R$ are negative and positive, respectively. 
The magnitudes of the polarizabilities in the vicinities of $\bar{\lambda}_{B}$  and $\bar{\lambda}_R$ are on the order of $-600$~a.u. and 3000~a.u., respectively. 
The magnitudes of the polarizabilities in the vicinity of $\bar{\lambda}_{B}$ are about five times smaller than in the vicinity of 
$\bar{\lambda}_R$. The reason is that $\bar{\lambda}_{B}$ is farther from 
the $D_1$ and $D_2$ resonance lines than $\bar{\lambda}_R$.

\subsection{Light shifts due to a linearly polarized light}

We calculate the light shifts of the transitions from the $6P_{3/2}FM_F$ sublevels to 
the $6S_{1/2}F^\prime M_F^\prime$ sublevels. 
The light shift of an atomic transition is the difference between the light shifts 
of the upper and lower levels. The light shifts of the hfs sublevels $|FM_F\rangle$ of the 
$6P_{3/2}$ excited state  
are determined by diagonalizing the Hamiltonian (\ref{8}), 
which includes the hfs energy (\ref{6}) and the Stark interaction energy (\ref{3b}).
The Stark energy of the $6P_{3/2}$ state is produced by the scalar polarizability $\alpha_0(6P_{3/2})$ and the tensor
polarizability $\alpha_2(6P_{3/2})$. All the hfs sublevels $|F^\prime M_F^\prime\rangle$ of the $6S_{1/2}$ ground state 
have the same light shift, 
produced by the scalar polarizability $\alpha_0(6S_{1/2})$.
For simplicity, we assume in this subsection that the electric component of the light field is 
linearly polarized along the quantization axis $z$. In addition, we limit ourselves to the transitions from the excited-state sublevels that are split from $F=5$. 
  
\begin{figure}
\begin{center}
  \includegraphics{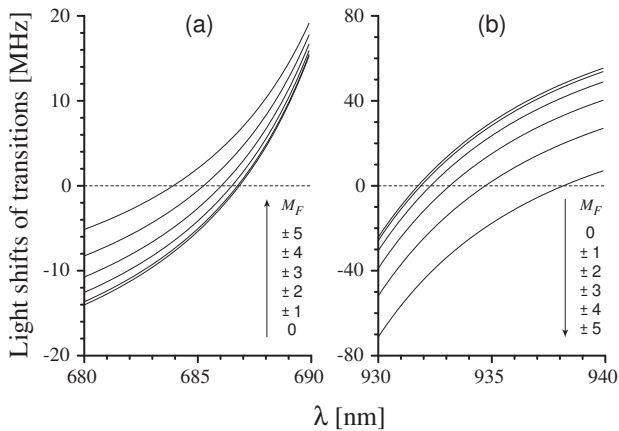}
 \end{center}
\caption{Light shifts of the transitions from the $6P_{3/2}FM_F$ sublevels to the 
$6S_{1/2}F^\prime M_F^\prime$ sublevels in cesium as functions of the light wavelength in the vicinities of the central blue-detuned magic wavelength (a) and the central red-detuned magic wavelength (b). The electric component of the field is linearly polarized along the quantization axis $z$. The intensity of the field is 1~MW/cm$^2$. We show only the results for the transitions involving the excited-state sublevels that are split from $F=5$.}
\label{fig5}
\end{figure}

In Fig. \ref{fig5}, we plot the shifts of the transition frequencies as functions of the light wavelength in the vicinities of the central blue-detuned magic wavelength $\bar{\lambda}_{B}=685.5$ nm (a) and the central red-detuned magic wavelength $\bar{\lambda}_R=934.5$ nm (b). 
As seen,  the light shifts cross zero at around $\bar{\lambda}_{B}$ and $\bar{\lambda}_R$, with  positive slopes. 
At $\bar{\lambda}_{B}$  and $\bar{\lambda}_R$, the light shifts range from $-4$ MHz to 
$3.1$ MHz and from $-21.2$ MHz to $26.3$ MHz, respectively. 
The range of the light shifts in the vicinity of $\bar{\lambda}_{B}$ is several times smaller than that in the vicinity of $\bar{\lambda}_R$. This is due to the difference between the 
magnitudes of the polarizabilities around $\bar{\lambda}_{B}$ and $\bar{\lambda}_R$.

\begin{figure}
\begin{center}
  \includegraphics{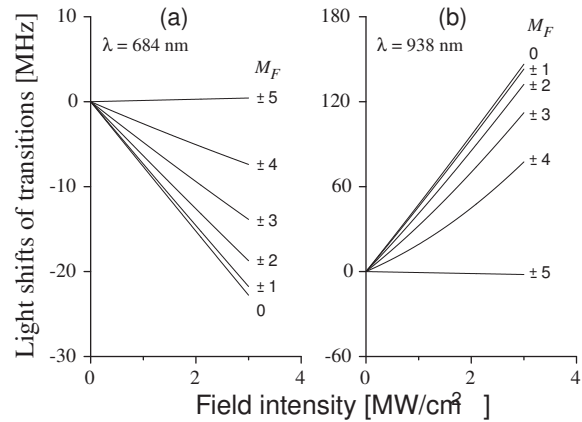}
 \end{center}
\caption{Light shifts of the transitions from the $6P_{3/2}FM_F$ sublevels to the 
$6S_{1/2}F^\prime M_F^\prime$ sublevels in cesium vs. the field intensity. The values of the wavelength of the field are chosen to be $\lambda=684$ nm (a) and $\lambda=938$ nm (b). The electric component of the field is linearly polarized along the quantization axis $z$. We show only the results for the transitions involving the excited-state sublevels that are split from $F=5$.}
\label{fig6}
\end{figure}

In Fig. \ref{fig6}, we plot the shifts of the transition frequencies as functions of the field
intensity. We choose the wavelengths $\lambda=684$ nm (a) and $\lambda=938$ nm (b) for the field. These values are close to the exact values of the blue- and red-detuned magic wavelengths for the transitions $6P_{3/2}F=5\,M_F=\pm5\leftrightarrow 6S_{1/2}F^\prime\,M_F^\prime$ (with
the maximum values of $F$ and $|M_F|$). 
Our numerical calculations show that the light shifts of 
these transitions are indeed small. 
They are less than 2 MHz even when the field intensity is as high as 3 MW/cm$^2$. A more precise tuning can, in principle, 
reduce the light shifts of these transitions to zero. The light shifts of the other transitions 
are more substantial but not very large. At an intensity of  3 MW/cm$^2$, the light shifts of the 
transitions from the $F=5\,M_F=0$ sublevel are  $-23$ MHz in the case of Fig. \ref{fig6}(a)
and 146 MHz in the case of Fig. \ref{fig6}(b). 
In general, it is possible to individually minimize the light shifts of the transitions from the upper sublevels $6P_{3/2}FM_F$ with arbitrary fixed values of $F$ and $M_F$  using appropriate choices of magic wavelengths.

\section{Light shifts of the transitions of atomic cesium in a two-color evanescent field around a subwavelength-diameter fiber}
\label{sec:Stark}

Consider a cesium atom moving outside  a thin  single-mode optical fiber that has a cylindrical silica core of radius $a$ and refractive index $n_1$ and  an infinite vacuum clad of refractive index $n_2=1$. To produce an optical potential with a trapping minimum sufficiently far from the fiber surface, we use two laser beams propagating along the fiber   
in the fundamental modes 1 and 2 with differing frequencies $\omega_1$ and $\omega_2$, respectively (with wavelengths $\lambda_1$ and $\lambda_2$, respectively, and free-space wave numbers $k_1$ and $k_2$, respectively) \cite{twocolors}. To make the potential cylindrically symmetric, the laser beams are circularly polarized at the input. 
In the vicinity of the fiber surface, the polarization of the transverse component of each propagating field rotates elliptically in time, 
the orbit  rotates circularly in space, and the spatial  distribution of the field intensity is cylindrically symmetric  \cite{paper 2}. 
For certainty, we assume that circulation of photons around the fiber axis $z$ is clockwise. 

Outside the fiber, in the cylindrical coordinates $\{r,\varphi,z\}$, 
the cylindrical components of the envelope vector $\vec{\mathcal{E}}$ of the electric field in a fundamental mode  with clockwisely rotating (circulating) polarization are given  
by \cite{fiber books}
\begin{eqnarray}
\mathcal{E}_r&=&iA [(1-s)K_0(qr)+(1+s)K_2(qr) ]
e^{i(\beta z- \varphi)},
\nonumber\\
\mathcal{E}_\varphi&=&A [(1-s)K_0(qr)-(1+s)K_2(qr) ]
e^{i(\beta z- \varphi)},
\nonumber\\
\mathcal{E}_z&=& A\frac{2q}{\beta}K_1(qr)e^{i(\beta z- \varphi)}.
\label{10}
\end{eqnarray} 
Here $\beta$ is the longitudinal propagation constant determined by the eigenvalue equation 
for the fiber mode with the free-space wavenumber $k=\omega/c$, 
the parameter $q=(\beta^2-n_2^2k^2)^{1/2}$ characterizes the decay of the field outside the fiber, and the parameter $s$ is defined as 
$s=({1}/{q^2a^2}+{1}/{h^2a^2})/[{J_1^\prime (ha)}/{haJ_1(ha)}+{K_1^\prime (qa)}/{qaK_1(qa)}]$, 
with $h=(n_1^2k^2-\beta^2)^{1/2}$.
The coefficient $A$ is proportional to the amplitude of the field. 
The notation $J_n$ and $K_n$ stand for the  Bessel functions of the first kind and the modified Bessel functions of the second kind, respectively.

\begin{figure}
\begin{center}
  \includegraphics{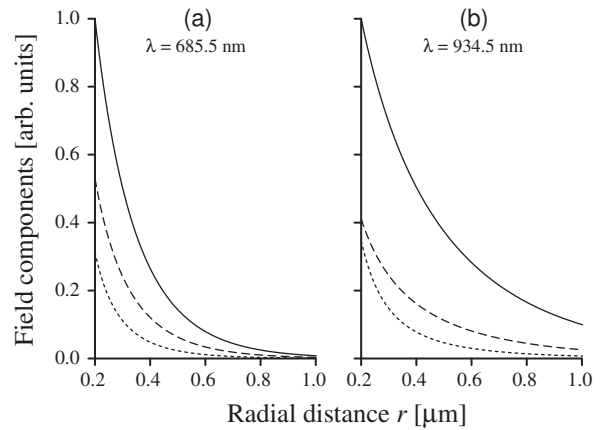}
 \end{center}
\caption{Magnitudes $|\mathcal{E}_{-1}|$ (dotted lines), $|\mathcal{E}_0|$ (dashed lines), and $|\mathcal{E}_1|$ (solid lines) of the spherical tensor components  of the evanescent light field in a fundamental mode with clockwisely rotating polarization outside a vacuum-clad subwavelength-diameter fiber. The radius of the fiber is $a=0.2$ $\mu$m. 
The light wavelength is $\lambda=685.5$ nm (a) and $934.5$ nm (b).
} 
\label{fig7}
\end{figure}

In the spherical tensor representation, the components of the field envelope are given by
\begin{eqnarray}
\mathcal{E}_{-1}&=&\sqrt{2}\,iA (1+s)K_2(qr) 
e^{i(\beta z- 2\varphi)},
\nonumber\\
\mathcal{E}_0&=& A\frac{2q}{\beta}K_1(qr)e^{i(\beta z- \varphi)},
\nonumber\\
\mathcal{E}_1&=&-\sqrt{2}\,iA (1-s)K_0(qr)e^{i\beta z}.
\label{11}
\end{eqnarray}
In the case of conventional weakly guiding fibers \cite{fiberbooks},  $\mathcal{E}_{-1}$ and $\mathcal{E}_{0}$ are negligible compared to $\mathcal{E}_{1}$. However, in the case of vacuum-clad subwavelength-diameter fibers, $\mathcal{E}_{-1}$ and $\mathcal{E}_{0}$ are not negligible \cite{paper2}. 

We illustrate in Fig.~\ref{fig7} the magnitudes $|\mathcal{E}_{-1}|$, $|\mathcal{E}_{0}|$, and $|\mathcal{E}_{1}|$ of the spherical tensor components 
of the field outside the fiber. According to the figure, all the three components of the field are comparable to each other in the vicinity of the fiber surface. Therefore, we must take into account all of these terms when we calculate the light shifts of the excited-state levels of atoms outside the thin fiber.

We use eqs. (\ref{3b}) and (\ref{11}) together with the Hamiltonian (\ref{8}) and the hfs energy
(\ref{6}) to calculate the light shifts of the $6P_{3/2}$ hfs sublevels of atomic cesium in a two-color evanescent field around a vacuum-clad subwavelength-diameter fiber. 
These light shifts are in fact equal to the optical potentials of the excited atoms outside the fiber. In addition, we also calculate the light shift of the $6S_{1/2}$ state, which is
the optical potential of the ground-state atoms.  
To minimize the difference between the light shifts of the $6P_{3/2}$  and 
$6S_{1/2}$ states, we choose the central red- and blue-detuned magic wavelengths for the two fields, i.e., $\lambda_1=\bar{\lambda}_R=934.5$ nm  and $\lambda_2=\bar{\lambda}_B=685.5$ nm.  
The powers of the laser beams must be chosen appropriately so that the ground-state optical potential has a deep minimum outside the fiber with a high barrier near the fiber surface. 
For this purpose, we choose the powers $P_1=11.5$ mW and $P_2=48.5$ mW
for the red- and blue-detuned laser beams, respectively. 

\begin{figure}
\begin{center}
  \includegraphics{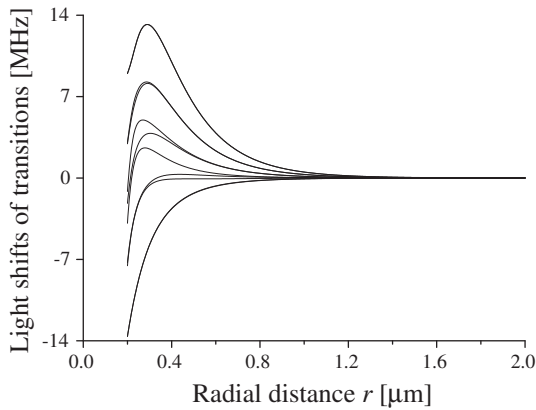}
 \end{center}
\caption{Spatial dependences of the light shifts of the $6P_{3/2}FM_F\leftrightarrow 6S_{1/2}F^\prime M^\prime_F$  transitions of cesium atoms in a two-color evanescent field around a vacuum-clad silica-core subwavelength-diameter fiber. 
The radius of the fiber is $a=0.2$ $\mu$m. 
The two laser beams are tuned to the wavelengths $\lambda_1=\bar{\lambda}_R=934.5$ nm  and $\lambda_2=\bar{\lambda}_B=685.5$ nm, with the powers $P_1=11.5$ mW and $P_2=48.5$ mW, respectively. 
We show only the results for the transitions involving the excited-state sublevels that are split from $F=5$.
} 
\label{fig8}
\end{figure}

We plot in Fig. \ref{fig8} the light shifts of the $6P_{3/2}FM_F\leftrightarrow 6S_{1/2}F^\prime M^\prime_F$  transitions of the atoms trapped around the fiber.
The figure shows that the light shifts can be reduced to become less than 14 MHz. Such a shift is comparable to the typical detuning of near-resonant fields used for a MOT with cesium atoms (typical detuning is 10--20 times of the natural linewidth 
$\gamma=5.18$ MHz of the cesium $D_2$ line) \cite{dipoleforce}. 
Thus, due to the use of the red- and blue-magic wavelengths, 
the spatial dependences of the light shifts of atomic transitions are weak. This opens up an opportunity for state-insensitive trapping, i.e., for simultaneous trapping of ground- and excited-state atoms  \cite{Katori,Kimble}. If we detune the MOT fields from the cooling transition  
by a negative detuning $\delta$ with a magnitude $|\delta|/2\pi>14$ MHz, then the red-detuning condition for the MOT fields (cooling fields) in the presence of the far-off-resonance evanescent fields (optical trapping fields) is kept throughout the outside of the fiber. 
This allows the simultaneous operation of the MOT and the dipole trap. 

\begin{figure}
\begin{center}
  \includegraphics{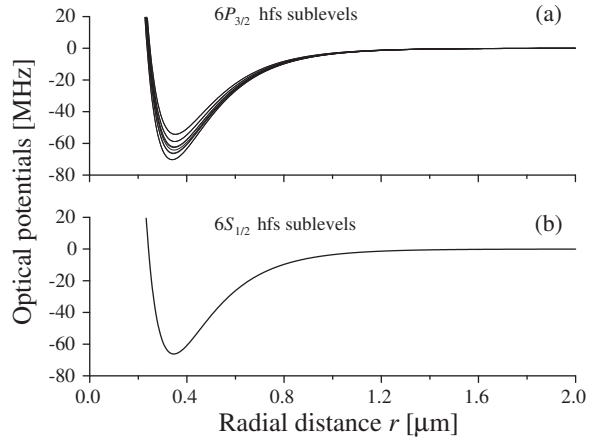}
 \end{center}
\caption{Optical potentials (light shifts) of  cesium atoms in a two-color evanescent field around a vacuum-clad subwavelength-diameter fiber. 
The atoms are in the excited-state manifold $6P_{3/2}$ (a) 
and the ground-state manifold $6S_{1/2}$ (b). 
All the parameters are the same as those for Fig. \ref{fig8}. 
We show only the results for the transitions involving the excited-state sublevels that are split from $F=5$.
} 
\label{fig9}
\end{figure}

We plot  in Fig.~\ref{fig9} the optical potentials of the excited- and ground-state atoms for the parameters of Fig.~\ref{fig8}. 
We observe from Fig.~\ref{fig9}
that the excited- and ground-state optical potentials have similar shapes, 
with deep minima located close to each other in space. This indicates the possibility
of state-insensitive two-color trapping of cesium atoms around the fiber. 
Such a state-independent trapping scheme allows the simultaneous operation of trapping and
probing, that is, the operation of trapping with continuous observation during the trapping interval.
Kimble \textit{et al.} have proposed and demonstrated 
a similar state-insensitive trapping method, which is based on the use of counter-propagating laser beams at a red-detuned magic wavelength \cite{Kimble}.

\section{Conclusions}
\label{sec:summary}
In summary, we have shown that the light shifts of the $6P_{3/2}FM_F\leftrightarrow 6S_{1/2}F^\prime M_F^\prime$ transitions in atomic cesium can be minimized by tuning one trapping light to around  $934.5$ nm in wavelength (central red-detuned magic wavelength) and the other light to around $685.5$ nm in wavelength (central blue-detuned magic wavelength).
We have investigated the light shifts of the cesium hfs sublevels  in a two-color evanescent field around a subwavelength-diameter fiber. 
The simultaneous use of the red- and blue-detuned magic wavelengths allows state-insensitive two-color trapping and guiding of cesium atoms along the thin fiber.
Our results can be used to efficiently load a two-color dipole trap by cesium atoms from
a magneto-optical trap and to perform continuous observations. 

\section*{Acknowledgment}

This work was carried out under the 21st Century COE program on ``Coherent Optical Science''.


\begin{thebibliography}{99}

\bibitem{Schlosser} N. Schlosser, G. Reymond, I. Protsenko and P. Grangier:
Nature  \textbf{411} (2001) 1024.

\bibitem{Kuhr} S. Kuhr, W. Alt, D. Schrader, M. M\"{u}ller, V. Gomer and D. Meschede: 
Science \textbf{293} (2001) 278. 

\bibitem{Sackett} C. A. Sackett, D. Kielpinski, B. E. King, C. Langer, V. Meyer, C. J. Myatt, M. Rowe, Q. A. Turchette, W. M. Itano, D. J. Wineland and C. Monroe: Nature  \textbf{404} (2000) 256.

\bibitem{dipoleforce} A. P. Kazantsev, G. J. Surdutovich and V. P. Yakovlev: \textit{Mechanical Action of Light on Atoms} (World Scientific, Singapore, 1990); R. Grimm, M. Weidem\"{u}ller and Yu. B. Ovchinnikov: Adv. At., Mol., Opt. Phys. \textbf{42} (2000) 95; V. I. Balykin, V. G. Minogin and V. S. Letokhov: Rep. Prog. Phys. \textbf{63} (2000) 1429. 

\bibitem{paper1} 
V. I. Balykin, K. Hakuta, Fam Le Kien, J. Q. Liang and  M. Morinaga:  Phys. Rev. A \textbf{70} (2004) 011401(R); 
V. I. Balykin, Fam Le Kien, J. Q. Liang,  M. Morinaga and K. Hakuta: \textit{CLEO/IQEC and PhAST Technical Digest} on CD-ROM (Optical Society of America, Washington, DC 2004), presentation ITuA7.

\bibitem{twocolors} Fam Le Kien, V. I. Balykin and K. Hakuta: Phys. Rev. A \textbf{70} (2004) (accepted).

\bibitem{Ovchinnikov} Yu. B. Ovchinnikov, S. V. Shul'ga and V. I. Balykin: J. Phys. B \textbf{24} (1991) 3173.

\bibitem{Domokos} P. Domokos, P. Horak and H. Ritsch: Phys. Rev. A \textbf{65} (2002) 033832.

\bibitem{coolingbook} H. J. Metcalf and P. van der Straten: \textit{Laser Cooling and Trapping} (Springer, New York, 1999).

\bibitem{Kuppens} S. J. M. Kuppens, K. L. Corwin, K. W. Miller, T. E. Chupp and C. E. Wieman: Phys. Rev. A \textbf{62} (2000) 013406.

\bibitem{Katori} H. Katori, T. Ido and M. Kuwata-Gonokami: J. Phys. Soc. Jpn. \textbf{68}, (1999) 2479; T.  Ido, Y. Isoya and H. Katori: Phys. Rev. A \textbf{61} (2000) 061403(R).

\bibitem{Kimble} J. McKeever, J. R. Buck, A. D. Boozer, A. Kuzmich, H.-C. N\"{a}gerl, D. M. Stamper-Kurn and
H. J. Kimble: Phys. Rev. Lett. \textbf{90} (2003) 133602.

\bibitem{Mazur'sNature}  L. Tong, R. R. Gattass, J. B. Ashcom, S. He, J. Lou, M. Shen, I. Maxwell and  E. Mazur: Nature \textbf{426} (2003) 816.

\bibitem{Birks} T. A. Birks, W. J. Wadsworth and P. St. J. Russell:  Opt. Lett. \textbf{25} (2000) 1415;
S. G. Leon-Saval, T. A. Birks, W. J. Wadsworth, P. St. J. Russell and  M. W. Mason:
\textit{Conference on Lasers and Electro-Optics (CLEO)},
Technical Digest, Postconference Edition (Optical Society of America, Washington, DC 2004),
paper CPDA6. 

\bibitem{Schwartz} C. Schwartz:  Phys. Rev. \textbf{97} (1955) 380.

\bibitem{Schmieder} R. W. Schmieder: Am. J. Phys. \textbf{40} (1972) 297.

\bibitem{Khadjavi} A.  Khadjavi, A. Lurio and W. Happer:  Phys. Rev.  \textbf{167} (1968) 128.

\bibitem{Schmieder71} R. W. Schmieder, A. Lurio and W. Happer:  Phys. Rev. A  \textbf{3} (1971) 1209.

\bibitem{Boyd} See, for example, R. W. Boyd: \textit{Nonlinear Optics} (Academic, New York, 1992). 

\bibitem{Safronova} M. S. Safronova and C. W. Clark:  Phys. Rev. A \textbf{69} (2004) 040501(R) and references therein.

\bibitem{Fabry} M. Fabry and J. R. Cussenot: Can. J. Phys. \textbf{54} (1976) 836.

\bibitem{Moore} C. E. Moore: \textit{Atomic Energy Levels}, Natl. Bur. Stand. Ref. Data Ser. Natl. Bur. Stand. (U.S.) Circ. No. 467 (U.S. GPO, Washington, D.C., 1971), Vol. 35.

\bibitem{Theodosiou} C. E.  Theodosiou: Phys. Rev. A  \textbf{30} (1984) 2881.

\bibitem{paper2} Fam Le Kien, J. Q. Liang, K. Hakuta and V. I. Balykin: 
Opt. Commun. \textbf{} (2004) (in press).

\bibitem{fiberbooks} See, for example, 
D. Marcuse: \textit{Light Transmission Optics} 
(Krieger, Malabar, FL,  1989).

\end{thebibliography}
\end{document}